\documentclass[twocolumn]{mn2e}
\usepackage{epsfig}
\usepackage{epstopdf}
\usepackage{subfloat}
\usepackage{subfigure}
\usepackage{hyperref}
\usepackage{multirow}
\usepackage{graphicx,times}  
\usepackage{soul}
\usepackage{natbib}
\usepackage{amssymb,amsmath}
\usepackage{float}

\def\HI{{H~{\sc i} }}
\def\galaxy{NGC~3621 }
\begin{document}

\title{Bending Waves in Velocity Space: a First Look at the THINGS sample}
%\title{Velocity corrugations - signature of bending waves in face on disc galaxies?}
\author[]
{Meera Nandakumar$^{1}$\thanks{meeranandakr.rs.phy17@itbhu.ac.in}, 
Chaitra Narayan$^{2}$\thanks{chaitra@ncra.tifr.res.in},
Prasun Dutta$^{1}$\thanks{pdutta.phy@itbhu.ac.in},  
%Jayaram Chengalur$^{2}$
\\$^{1}$Department of Physics, IIT (BHU), Varanasi 221005, India. 
\\$^{2}$National Centre for Radio Astrophysics, Tata Institute of Fundamental Research, Pune University, Pune 411007, India.
}
\maketitle 

\begin{abstract}
Detection of bending waves is a highly challenging task even in nearby disc galaxies due to their sub-kpc bending amplitudes. However, simulations show that the harmonic bending of a Milky Way like disc galaxy is associated with a harmonic fluctuation in the measured line of sight (los) velocities as well, and can be regarded as a kinematic signature of a manifested bending wave. Here, we look for similar kinematic signatures of bending waves in  \HI discs, as they extend to much beyond the optical radii. 
	We present a multipole analysis of the \HI los residual velocity fields of six nearby spiral galaxies from the THINGS sample, which uncovers the bending wave-induced velocity peaks.  This allows us to identify the radial positions and amplitudes of the different bending modes present in the galaxies. We find that all of our sample discs show a combined kinematic signature of superposition of a few lower-order bending modes, suggesting that bending waves are a common phenomenon. The identified velocity peaks are found to be of modes $m=2,3$ and $4$, not more than 15 km s$^{-1}$ in amplitude and spread across the entire \HI disc.  Interestingly, they appear to be concentrated near the optical edge of their host galaxies. Also, $m=2$ appears to be more common than the other two modes.
\end{abstract}


\section{Introduction}
\label{sec:Intro}

A large number of observers have noted local wavy undulations of the disc components in our Galaxy \citep{Gum+1960,Dixon1967,Lynga1970,Quiroga1974,Lockman1977,CohenThaddeus1977,Spicker1986,Alves+2020,Thulasidharan+2022}.
 These `local' vertical deviations of the mid-plane of about 70 - 100 pc, initially termed as `vertical corrugations' \citep{Quiroga1974} or `scalloping' \citep{Kulkarni+1982}, were noted to differ from warps in size, shape and location in the Milky Way disc.
Subsequently, theories were proposed \citep{Lynden-Bell1965, HunterToomre1969, Nelson1976,Shu+1983} that regarded both warps and corrugations as `Bending Waves' - the global bending of a disc in response to small non-axisymmetric perturbations. 
We refer the reader to several comprehensive introductions \citep{Sanchez+2015, ChequersWidrow2017, Laporte+2018a, Poggio+2021, Gomez+2021} that do justice to this decades-old topic. 
In short, the Bending Wave theory can successfully explain the observed shapes of disc bending with a single parameter - the azimuthal bending mode, m. 
 The mode m$=$0 can lead to the U-shape (bowl mode) seen in NGC 4650A \citep{Sparke1995, Whitmore1991} and NGC 4631 \citep{Richter+2018}; m$=$1 describes the most commonly observed S-shaped warps; m$=$2 can give rise to a saddle shape, and corrugations could arise from a large m value (like m$>$ 10).

Although the global presence of m = 0, 1 and 2 in the Milky Way's \HI disc were discovered by \cite{Levine+2006}, it was only after the advent of stellar kinematic surveys covering large parts of the Galaxy, like RAVE (Radial Velocity Experiment), GAIA, SDSS (Sloan Digital Sky Survey), LAMOST (Large Sky Area Multi-Object Fibre Spectroscopic Telescope) etc., that the global nature of disc bending (for m $\neq$ 1) received due attention. 
In a series of discoveries, substructures related to bending and breathing modes have since been identified \citep{Widrow+2012, Gomez+2012a, Williams+2013, YannyGardner2013, Slater+2014, Price-Whelan+2015, Xu+2015, Morganson+2016, Schoenrich&Dehnen2018, Wang+2018,BennettBovy2019,Carrillo+2019,Cheng+2020,Wang+2020,Lopez+2020} from within the solar radius up to the edges of the stellar disc.
In addition, the presence of the phase spiral \citep{Antoja+2018} was also revealed.
Using a toy model, \cite{BS18} show that the phase spiral can be the result of a co-evolution of both density wave and bending wave generated by the crossing of Sagittarius Dwarf through the Galactic disc about 0.5 Gyr ago.
These discoveries fuelled a frenzy of parallelly evolving simulations \citep{Gomez+2013, Gomez+2016, Gomez+2017, Gomez+2021, Bland-Hawthorn+2019, BT21, Laporte+2019, ChequersWidrow2017, Chequers+2018, Poggio+2021, GrionFilho+2021, Hunt+2021, Schoenrich&Dehnen2018, Widrow+2014, Khoperskov+2019} that were able to reproduce these substructures as due to bending waves generated under different scenarios. 

A number of physical mechanisms have been put forward to generate the bending waves in disc galaxies, such as tidal interaction with satellites or companion galaxies \citep{HunterToomre1969, Edelsohn&Elmegreen1997, Weinberg1998, Schwarzkopf&Dettmar2001,WeinbergBlitz2006, Gomez+2013, Widrow+2014, Gomez+2016, Gomez+2017, Schoenrich&Dehnen2018, Bland-Hawthorn+2019, Laporte+2019, BT21, Gomez+2021, Poggio+2021, GrionFilho+2021, Hunt+2021}; intergalactic matter accreted onto the dark halo \citep{JiangBinney1999} or directly onto the disc \citep{Lopez-Corredoira+2002}; intergalactic magnetic field \citep{Battaner+1990} and intergalactic wind \citep{KahnWoltjer1959}. 
Bending waves are shown to also arise, from various internal instabilities \citep{Araki1985,RevazPfenniger2004,Sellwood1996,ChequersWidrow2017}, from resonant coupling \citep{Binney1981}, due to the dynamical friction between a disc and its halo \citep{NelsonTremaine1995} and even due to the halo substructure \citep{Chequers+2018}. 
While the number of mechanisms put forth to generate bending waves helps us understand the observed frequency of warps \citep{ReshComb1998,AnnPark2006}, it also raises the curiosity about the rest of the bending modes (m$\neq$1) that are not so visible in disc galaxies? 

Recent simulations of Milky-way like galaxies by \cite{Gomez+2017, ChequersWidrow2017} show that corrugations must be as common as warps with quite a high chance of developing fairly long-lasting complicated vertical structures. \cite{ChequersWidrow2017} also compile a list of ‘minimum requirements’ needed to observe bending waves in external galaxies.
Observationally though, the higher-order modes are seen in not more than 10 galaxies including the Milky Way \citep{Narayan+2020}. 
This may be due to their low bending amplitudes - from the nine instances of measured corrugation amplitudes available so far (see Table 2 in \cite{Narayan+2020}), none are found to exceed 300 pc. 
Such a small magnitude of bending can be hidden within the disc thickness, and go undetected. 
The typical bending amplitude in warps on the other hand is $\sim$ $1$ kpc (often larger in \HI discs) and are thus easily noticed.
Further, all corrugations measured so far are in edge-on galaxies.  As a corrugated disc tilts towards being face-on, the visible bending goes down by a factor of sin(i) \citep{Fridman+1998}, where i is the inclination angle and detection becomes difficult.

While direct detection of corrugation in neighbouring discs seems far-fetched and limited to near edge-on discs, several authors have looked for possible kinematic signatures (wavy undulations in mean velocity perpendicular to disc) in face-on discs \citep{Alfaro+2001, Sanchez+2015, Donghia+2016,Gomez+2021} and even found them. 
Yet the problem is that spatial corrugation and its associated vertical velocity corrugation cannot be measured (with the exception of Milky Way) in the same system.
Therefore the observed velocity corrugation could not have been attributed unambiguously to disc bending.
This issue was resolved after a series of recent simulations \citep{Gomez+2013, Gomez+2016, Gomez+2017, Gomez+2021, Bland-Hawthorn+2019, BT21, Laporte+2019, Widrow+2014, ChequersWidrow2017, Chequers+2018} studying the bending wave generation and evolution in Milky Way - Sagittarius Dwarf interaction, found that the bending waves are always associated with an oscillatory pattern in the disc's mean vertical velocity. In these studies, they also establish a quantitative connection between bending waves and vertical velocity patterns. 
\cite{Darling&Widrow2019b} find that a mid-plane deviation of 2kpc can lead to a change in mean vertical velocity of about $50-60$ km s$^{-1}$ in their simulations of the Milky Way.
\cite{Antoja+2018} observe a similar connection in our Galaxy where a mid-plane bend of 1kpc corresponds to $30$ km s$^{-1}$.
\cite{Laporte+2018b} note that the velocity fluctuations can vary from almost none in the inner galaxy to $\sim$10 km s$^{-1}$ at R $=10$kpc.
In the outer disc, beyond the Monoceros ring, these fluctuations can reach as much as 50 km s$^{-1}$.
 
For galaxies that are not edge-on, detection of modes in vertical velocity components can imply the presence of bending waves. This opens up a new avenue towards quantifying the depth of a tidal interaction \citep{Poggio+2021, GrionFilho+2021} for a large population of inclined galaxies. 
We could also use this proxy method to address the pressing questions in the hitherto less explored bending wave family (m $\neq$ 1) such as (i) are they as common as warps? (ii) how long are they stable? and (iii) what is their impact on galaxy evolution? This motivates us to look for kinematic imprints of bending waves in the vast \HI database of nearby galaxies. Galaxies from {\bf T}he {\bf HI N}earby {\bf G}alaxy {\bf S}urvey (THINGS henceforth)  \citep{Walter+2008} serves as a good starting point as it provides the spatial and velocity resolution required to identify the velocity undulations. 

It has been customary to model the non-axisymmetric features of a disc galaxy from its line of sight (los) velocity maps using the framework laid out by \cite{Schoenmakers+1997}. In fact, this method has been used by \cite{Trachternach+2008} and \cite{Schmidt+2016} on the THINGS sample too. 
In this method, the los velocity data is subject to Fourier decomposition in order to extract the non-axisymmetric features with the key assumption that the entire contribution to the los velocity comes from in-plane motion of the gas, i.e., circular and radial. The disc is assumed to be flat without any vertical bending. These assumptions are also the basis of routines applied to los velocity maps (For eg, VELFIT by \cite{Spekkens&Sellwood2007}).
For galaxies known to have warps, care is taken to limit the analysis to within warp starting radius \citep{Schoenmakers+1997, Trachternach+2008}. Consequently, any uncovered global azimuthal modes are attributed to in-plane non-axisymmetries like the spiral arms, non-spherical halo potential etc. In more recent works \cite{Sellwood&Spekkens2015, Schmidt+2016} have relaxed the flat disc assumption to accommodate for outer disc warps but even here the vertical velocity contribution is restricted to m$=$1 and m$>$1 would not be represented. The recent developments suggesting that bending modes ought to be more common and that they can also lead to harmonic oscillations in los velocity data, has been ignored. Vertical velocity contribution in the analysis of los velocity maps needs to be included.

From a harmonic decomposition of the velocity fields of 19 galaxies from the THINGS sample, \cite{Trachternach+2008} find that the dark matter halo potential beyond optical radii is very close to spherical. \cite{Elson+2011} find a similar result for NGC 2915. They also find an axisymmetric radial outflow of $\sim$ 5-17 km s$^{-1}$ which is inconsistent with the possible disc formation scenario in which gas flows in from the surrounding intergalactic medium. These evidences weaken the case for a non-spherical halo and mass inflow as possible origins for non-axisymmetry. On the other hand, bending waves that can be easily excited under many circumstances, is more likely to be the cause. Besides, bending waves are expected to grow in amplitude in the outer disc regions \citep{Gomez+2013, BT21, Poggio+2021}. Thus we can safely assume that the non-axisymmetric global modes uncovered in the outer disc region using extended \HI are most likely due to disc bending. It is entirely possible that the axisymmetric radial outflow in NGC 2915 could be due to an m$=$0 (U-shaped) bending mode at the disc's edge.

In this paper, we present a multipole analysis of the \HI los residual velocity fields of six nearby spiral galaxies from the THINGS sample. We detect a total of 25 independent global modes spread across different galactocentric radii and different multipoles (m = 2,3 and 4). The paper is organized in the following order.  We discuss how the bending waves can be detected by extracting regular oscillatory patterns in the column density and line of sight velocity of \HI in Section~\ref{sec2}.  In Section~\ref{sec3}, we describe our analysis with a sample of six galaxies and explain various interesting trends seen here. We discuss the implication of these results in Section~\ref{sec:disc} and conclude with a summary of this study in  Section~\ref{sec:conc}.

\section{Signatures of bending waves in \HI observation}
\label{sec2}
\subsection{Kinematic signature of bending waves}
\label{sec:kinm}

\begin{figure}
%\begin{center}
\hspace*{-0.5cm}
\includegraphics[width=.5\textwidth]{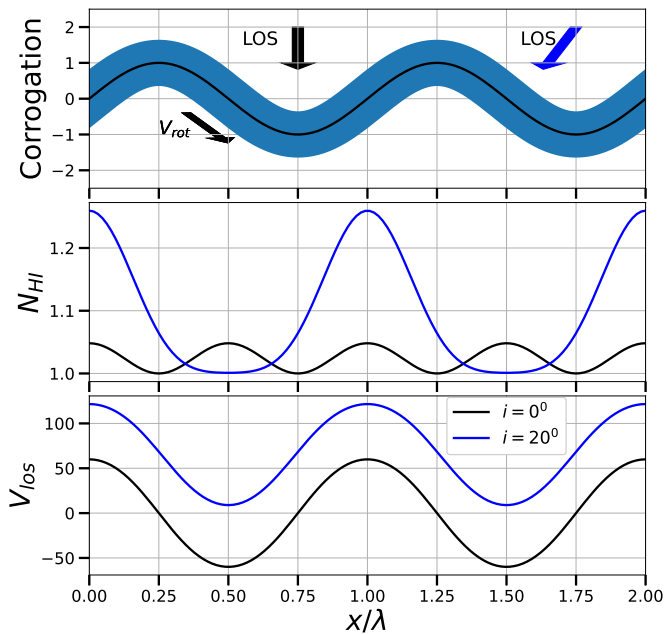}
\caption{A toy model showing  bending waves in a sheet  leading to column density and velocity corrugations. The top panel show the cross section of a corrugated disc with the $V_{rot}$ showing that the gas velocity follow the bending. The middle and bottom panel show the resulting column density and line of sight velocity oscillations (for $V_{rot} = 220$ km s$^{-1}$) for two inclination angles of $0^{\circ}$ (black) and $20^{\circ}$ (blue).  }
\label{fig:toymodel}
%\end{center}
\end{figure}  

Figure~\ref{fig:toymodel} shows how a bending wave in a disc can induce oscillation in los column density and line of sight velocity. The top panel shows the cross-section of a corrugated gas disc with a certain corrugation amplitude and thickness. 
Note that here the circular motion of the gas follows the corrugation pattern as shown by the black arrow with $V_{rot}$. We show the effect of observing such a disc from different inclination angles. From the top, that is from a face-on orientation, the motion of the gas and stars along the corrugated sheet produces an oscillatory line of sight velocity distribution. This is shown by the black line in the bottom panel. This oscillation has the same wavelength as that of the corrugation but is out of phase. The corresponding los column density variation is shown with the black curve in the middle panel. We notice, that if the disc is observed face-on the column density also shows an oscillatory pattern, however, the wavelength of oscillation, in this case, is half of that in the corrugation. The nature of the line of sight velocity and column density corrugation changes as the disc is viewed with higher inclination angles (blue curves). For an inclination of $20^{\circ}$ in our example, the amplitude of the line of sight velocity corrugation decreases and it picks up a constant component. At this inclination, the column density corrugation follows a periodic pattern that repeats at the same interval as the corrugation of the disc. For such a toy model, it can be trivially shown that for a corrugation with wavelength $\lambda$ and amplitude $h$, at inclination angles greater than $\tan^{-1} \left [ 2 \pi h / \lambda \right]$, the oscillations in column density and line of sight velocity are correlated. 

Local fluctuations of column density as well as velocity, may result also from compressive forces of self-gravity of the disc. \citet{2020MNRAS.496.1803N} estimate the column density and line of sight (los) velocity fluctuation power spectra for spiral galaxy NGC~5236. Their measured power spectra slopes suggest that fluctuations could be originated from compressive forces which can be from gravitational instabilities or self-gravity of the disc. It is then expected, that to maintain the vertical equilibrium of the disc, the regions with excess column density would also have an excess line of sight velocity dispersion. This would imply a correlation between the fluctuations in column density and fluctuations in los velocity dispersion. A similar trend is also expected in a corrugated disc, where the change in los path length crossing through the disc defines the fluctuations in both column density and velocity dispersion.  Hence, any correlation between the local fluctuation in the column density and line of sight velocity dispersion can imply either or both of the above reasons at work.

%\subsection{Quantifying oscillations in column density and line of sight velocity}
\subsection{Quantifying Bending modes using Harmonic Decomposition}
\label{sec:Meth}

%\begin{figure}
%\begin{center}
%\includegraphics[scale=.6]{6946_Corro.eps}
%\caption{The  fluctuations in column density  $\delta\N_{HI} (R, \phi)$, the line of sight velocity $\delta v (R, \phi)$ and line of sight velocity fluctuations $\delta \sigma_v (R, \phi)$ plotted as a function of $\phi$ for different radius $R$. The white horizontal line corresponds to the $r_{25}$ indicating the optical disc of the galaxy.}
%\label{fig:6946_Corro}
%\end{center}
%\end{figure}
Matter distribution and dynamics of the disc galaxies are often modelled in a cylindrical polar system with the origin at the centre of mass of the galaxy with coordinates $(R, \phi, z)$ with most of the matter being at the mid-plane $z=0$, hence a disc. Such an axisymmetric disc when subject to a non-axisymmetric perturbation can produce bending waves about the gravitational mid-plane such that at any given location ($R, \phi$), the vertical displacement of the disc mid-plane is $\Delta z \propto \cos{m \phi}$, $m$ being the azimuthal wave number. The wavelength of the bending waves should satisfy the $m \lambda = 2\pi R$ relation. It is worth noting here that \cite{Levine+2006} find that the observed bending in the Milky Way's \HI disc can be completely explained by the coexistence of $m = 0, 1$  and $2$ modes.

In this work, we look for the presence of similar lower-order bending modes in an external galaxy's \HI disc using the concept outlined in Sec \ref{sec:kinm}. For this, we investigate the fluctuations in the \HI column density, line of sight component of the velocity and velocity dispersion over their locally averaged values in the disc. 
%As the corrugations are vertical displacements as a function of $R$ and $\phi$, such fluctuations can arise as function of both of these coordinates. Here, we keep our interest in the corrugation on $\phi$ only and see it variation across different radius $R$. 
We use the zeroth, first and second-moment maps made using  \HI position-position-velocity data cubes obtained from radio interferometric observations of external spiral galaxies. For a typical galaxy, the disc may have an inclination with the line of sight of observation. Using information of the inclination and position angle from a tilted ring model fit to the rotation curve of the galaxy, we identify the coordinates  $R, \phi$ in the galaxy for each pixel in the moment maps. 
We choose a circular annulus of a radius, say, $R$ and width $\Delta R$.   The $\Delta R$ values are chosen to be equal for all radii in this analysis. The lower limit of $\Delta R$ then comes from the angular resolution. Since the galaxies have their disk oriented with the line of sight while observing different galacto-centric radii contribute to a given observation beam. Considering the local scale height of the galaxy be $h_z$ and the inclination angle is $i$, we choose $\Delta R < h_z\ \tan\, i$, to avoid any further mixing of contributions coming from different radii. We note that the \HI scale height of galaxies flares with higher scale height at a higher galactic centric radius \citep{1996AJ....112..457O,2016A&A...586A..98V}. The scale height of the galaxies in our sample is not estimated. Kregel et al. (2004) find that the ratio between the HI average scale height to the \HI scale length is $0.06 \pm 0.015$. We use the \HI extent $R_{HI}$ estimated as the semimajor axis value where the column density level of $10^{19}$ atoms/cm$^2$ to find a represented value for the scale height $h_z$ of the \HI disk. At the relatively outer galaxy, the actual scale height is expected to be higher and at the relatively lower radius, the scale height is expected to be lower. Hence we need to choose a much lower value than $h_z\ \tan\, i$, to avoid mixing of contributions from different radii. 
%{\bf  For a galaxy with inclination angle $i$ and typical disc scale height $h_z$, we keep $\Delta R < h_z  tan(i)$ to avoid blending of \HI coming from different $R$ due to projection. In addition we make sure that $\Delta R$ is sufficiently larger than synthesised beam (resolution) of the moment  maps.  }
We divide the  annulus in $N_{\phi}$ equal bins with each bin having an angular width of  $\delta \phi = 2\pi/ N_{\phi}$. For each bin, we estimate the bin value of the corresponding quantity as the average of all the pixel values from the moment map in the bin. For a typical annulus and a given moment map, this gives us the estimate of the azimuthal variation of the values at that radius $R$, we denote it by $f_j(R, \phi)$, where $j=0,1,2$ for column density, line of sight velocity and velocity dispersion respectively.  Uncertainties in these estimates, $\Delta f_j(R, \phi)$, are calculated by combining the measurement uncertainties in the pixels of the moment maps and the variation of the values in the pixels in a given bin. We do  harmonic decomposition of $f_j(R, \phi)$ to get the multipole amplitudes $A_m$ as follows:
\begin{equation}
\label{eq:f_phi}
    f_j(R, \phi)=\sum_{m=0}^{\infty} A_{mj}(R) cos(m\phi-\chi_{mj}(R)) \hspace{10pt}
\end{equation}
where $\chi_{mj} (R)$ is a phase factor.
Subsequently we will be interested in estimating the multipole amplitudes $A_{mj}(R)$ 
which can be obtained from 
\begin{equation}
\label{eq:Am}
A_{mj} = (\tilde{A}_{mj}^2 +\tilde{A^{'}}_{mj}^2)^{\frac{1}{2}}
\end{equation}
where $\tilde{A^{'}_{mj}}$ and $\tilde{A_{mj}}$ are defined as,
\begin{eqnarray}
\label{eq:Am_tilda}
\tilde{A}_{mj} = \frac{1}{2\pi}\int_0^{2\pi} f_j(R, \phi) cos(m\phi)d\phi \\ \nonumber
\tilde{A^{'}}_{mj}= \frac{1}{2\pi}\int_0^{2\pi} f_j(R, \phi) sin(m\phi)d\phi
\end{eqnarray}
%Note that the multipole $m=0$ corresponds to a azimuthally averaged monopole component, $m=1$ corresponds to a dipole etc.

We subtract the local averaged values of $f_j(R, \phi)$ from each of these bins to estimate the fluctuations in column density, line of sight velocity and velocity dispersion from the moment zero, one and two maps respectively. 
%Estimates of the local averaged values need to be different for different cases above and are discussed separately below.
\begin{itemize}
\item {\bf Column density:}
We consider the large scale distribution of the \HI column density in these galaxies are radial and estimate the fluctuations $\delta N_{HI} (R, \phi)$ on top of those. The radial profile at the radius $R$ significantly contributes to the  $A_{00}(R)$  multipole moment at that radius.
\item {\bf Line of sight velocity:}
Two major components in the line of sight velocity are from the systematic radial velocity of the galaxy and the rotation of the disc. The systematic radial velocity is estimated by the $A_{01}$ multipole amplitude, which gives the average line of sight velocity in the given annulus. Rotation of the disc induces a dipole component at a given radius with its value dependent on the tangential rotation velocity, inclination and position angle at that radius. This component is captured in the $m=1$ multipole mode $A_{11}(R)$.

\item {\bf Line of sight velocity dispersion:}
The azimuthally averaged \HI velocity dispersion is observed to reduce with $R$  and follow a steeper power-law with radius in the stellar disc compared to outside \citep{2009AJ....137.4424T}. We use $A_{02}(R)$ as an estimate of the azimuthally averaged line of sight velocity dispersion.  
\end{itemize}
In presence of bending modes of the disc, we expect to see oscillations in the column density and line of sight velocity as shown in Fig \ref{fig:toymodel}. We define the following quantities to measure them:
\begin{eqnarray}
\delta N_{HI} (R, \phi) &=& f_0(R, \phi) - A_{00}(R) \\  \nonumber 
\delta v (R, \phi) &=& f_1(R, \phi) - \left [A_{01} + A_{11}(R) cos(\phi + \chi_{11}(R)) \right ] \\ \nonumber
\delta \sigma_v (R, \phi)  &=& f_2(R, \phi) - A_{02}(R)
\end{eqnarray}
We estimate the spearman rank correlation between the values of $ \delta N_{HI} (R, \phi), \delta v (R, \phi)$ and $\delta \sigma_v (R, \phi)$ across different azimuth $\phi$ for a given $R$. 
As discussed earlier, the moment zero map of the  \HI position-position-velocity data cube gives the column density distribution. The higher moment maps are all weighted by the column density. Hence, it is possible that the column density fluctuations introduce fluctuations in higher-order moment maps and we should be cautious to interpret any nonzero $ \delta v (R, \phi)$ and $\delta \sigma_v (R, \phi)$ as fluctuations in the line of sight velocity or velocity dispersion. In such a case, however, the fluctuations in the line of sight velocity and its dispersion are expected to be correlated. Hence, if we observe a correlation between the line of sight velocity and its dispersion, we do not interpret the oscillations in velocity as arising from the corrugation in the disc. Furthermore, for a corrugated disc with a relatively higher inclination angle, the quantities $ \delta N_{HI} (R, \phi), \delta v (R, \phi)$ are expected to be correlated (or anti-correlated depending on the direction of the disc rotation velocity and position angle). On the other hand if the fluctuation in column density arises from self gravity, then we expect  $ \delta N_{HI} (R, \phi)$ and $\delta \sigma_v (R, \phi)$ to be correlated. Hence, looking for correlations among the residuals of the three-moment maps would be insightful.

%presence of these in the observed spearman rank coefficient gives important informations about the nature and origin of the corrugation.

%The cross correlation function between any two of the fluctuations $\delta A (R, \phi)$ and $\delta B(R, \phi)$ is defined as
%\begin{equation}
%\label{eq:crosscorr}
%\xi_{AB}(R, \psi) = \frac{ \int \limits_{0}^{2 \pi} \delta A (R, \phi) \delta B(R, \phi + \psi) d \phi } {\sqrt{ \int \limits_{0}^{2 \pi}[ \delta A (R, \phi) ]^2 d \phi \  \int \limits_{0}^{2 \pi}  \delta B(R, \phi)]^2  d \phi} } ,
%\end{equation}
%where $\psi$ is the lag in azimuth between the two quantities.

We use the multipoles  $m>0$ for column density and line of sight velocity dispersion and $m>1$
 for a line of sight velocity to seek possible evidence of corrugation in the disc.  We estimate the uncertainties in the multipole amplitude in the following way. We start by noting the uncertainties in $f_j(R, \phi)$, i.e  $\Delta f_j(R, \phi)$. Assuming the distribution of their uncertainties to be Gaussian random with mean  $f_j(R, \phi)$ and standard deviation $\Delta f_j(R, \phi)$, we generate several realisations of $f_j(R, \phi)$ and perform harmonic decomposition for each such realisation. The standard deviation of the amplitudes of these harmonic decompositions is then considered as the estimates of their uncertainties. 
 
 Note that, these multipole amplitudes will pick up the amplitude of a corrugation at the length scale of $2 \pi R / m$ present over the entire angular extent of the annulus. The multipole amplitudes calculated this way are however not sensitive to corrugations that may be present locally over smaller length scales. Hence, the analysis present in this section is not sensitive to any such small scale local corrugations.  

\section{Analysis and result}
\label{sec3}
\subsection{Sample Selection}
\label{sec:sample}

\cite{Walter+2008} observe \HI emission from a sample of 34 spiral galaxies for THINGS  using the VLA \footnote{VLA: Very Large array, NRAO, New Mexico} and estimate several \HI dynamical and morphological properties of these galaxies. They provide moment maps \footnote{Moment maps of THINGS galaxies are made publicly available in their %https://www2.mpia-hd.mpg.de/THINGS/Data.html
\href{https://www2.mpia-hd.mpg.de/THINGS/Data.html}{survey website}}
with angular resolutions of between $\sim 7 - 16^{''}$ and velocity resolutions $\sim 5$ km s$^{-1}$, though these numbers vary across the galaxies. 
%{\bf (See Table~\ref{tab:respar} for resolutions of different galaxies in parsec unit)}.
\cite{deBlok+2008} estimate the rotation curves of $19$ galaxies from the THINGS survey using tilted ring model fit, and also present the radial variation of disk inclination and position angles.  In this work, we investigate the regular pattern in the column density and los velocity that arises from an origin other than the large scale \HI variation over the radius and rotational velocity. Hence, we would need to include galaxies in our sample for which the rotation curves are estimated fairly well. For the relatively face-on galaxies, the rotational component of the velocity can not be well estimated, whereas for galaxies with high inclination, the line of sight passes through multiple radii and the sky position to galaxy position mapping is not clear.  Hence, we choose galaxies with an inclination angle between $30^{\circ}$ and $70^{\circ}$. We also restrict ourselves to the galaxies from THINGS with receding and approaching parts of the galaxies having nearly similar rotational velocities to avoid including discs with obvious warps in our sample. Keeping these in mind we choose six spiral galaxies from the THINGS sample for our analyses. Relevant properties of these galaxies are given in Table~\ref{tab:galpar}. The inclination and position angle given are disc averaged values for reference. We also list the optical radii $R_{25}$, where the B band surface brightness is at the level of $25$ magnitude arcsec$^{-2}$. The \HI extent $R_{HI}$  is defined as the semi-major axis at the level of column density of  $10^{19}$ atoms cm$^{-2}$.
 \begingroup
\begin{table}
	\caption{ Table giving  different properties of the six galaxies. Column 1: galaxy name; Column 2: distance (D), Column 3: average inclination angle (i); Column 4: average position angle (PA); Column 5: velocity resolution of THINGS data cube; Column 6:  \HI extent ($R_{HI}$) at a \HI column density of $10^{19}$ atoms cm$^{-2}$, Column 7: Optical extent ($R_{25}$),Column 8: References : (1) \citet{deBlok+2008}, (2) \citet{Walter+2008}, (3) \citet{2001ApJ...553...47F}, (4) \citet{2004AJ....127.2031K}, (5) \citet{2000A&AS..142..425D},(6) From LEDA survey \citep{2001misk.conf..683P},(7) \cite{2013NewA...19...89D}}
\setlength{\tabcolsep}{3.9pt} % Default value: 6pt
\renewcommand{\arraystretch}{.8}
\begin{tabular}{|l|c|c|c|c|c|c|l|}
\hline
Galaxy        & D     & i            & PA            & $\Delta v$      & $R_{HI}$ & $R_{25}$ & References \\ \hline
              & (Mpc) & ($^{\circ}$) & ($^{\circ}$) & (km sec$^{-1}$) & (kpc)    & (kpc)    &                   \\ \hline
NGC~2403 & 3.2   & 63           & 124          & 5.2             & 11.6     & 7.4      &   1,2,3,6,7         \\ \hline
NGC~2903 & 8.9   & 65           & 204          & 5.2             & 32.4     & 15.2     &   1,2,5,6,7         \\ \hline
NGC~3621 & 6.6   & 65           & 345          & 5.2             & 24.0     & 9.4      &   1,2,3,6,7        \\ \hline
NGC~4826 & 7.5   & 65           & 121          & 5.2             & 24.0        & 11.4     &   1,2,4,6,7         \\ \hline
NGC~5055 & 10.1  & 59           & 102          & 5.2             & 44.1     & 17.4     &   1,2,3,6,7         \\ \hline
NGC~6946 & 5.9   & 33           & 243          & 2.6             & 30.      & 9.8      &   1,2,3,6,7         \\ \hline
\end{tabular}
\label{tab:galpar}
\end{table}
\endgroup

\begin{figure}
\begin{center}
\includegraphics[scale=.25]{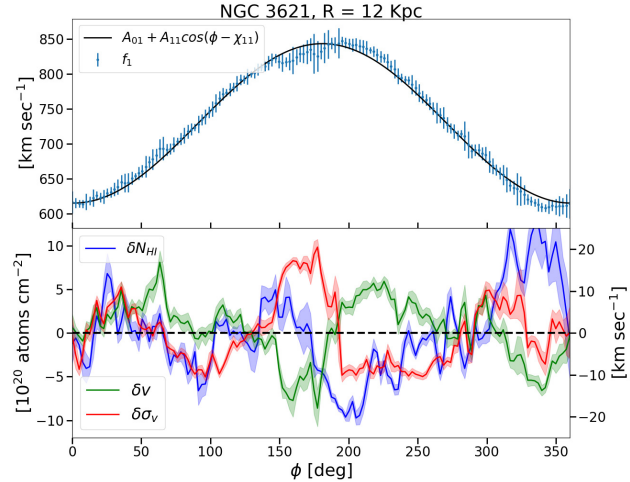}
\caption{ The top panel shows the function $f_1(R, \phi)$ for $R=12$ kpc (blue dots) for the galaxy \galaxy. Five-sigma uncertainties are also plotted for the same as error bars. The black continuous line represents the first two moments of this function. In the bottom panel we show the column density,  line of sight velocity and velocity fluctuations $ \delta N_{HI} (R, \phi), \delta v (R, \phi)$ and $\delta \sigma_v (R, \phi)$ with $\phi$ at the same radius $R$. The shaded regions represents one-sigma uncertainties. The y-axis on the left side shows column density scale and the y-axis on right side shows velocity scale.}
\label{fig:sing_rad}
\end{center}
\end{figure}

\begin{figure}
\begin{center}
\includegraphics[scale=.3]{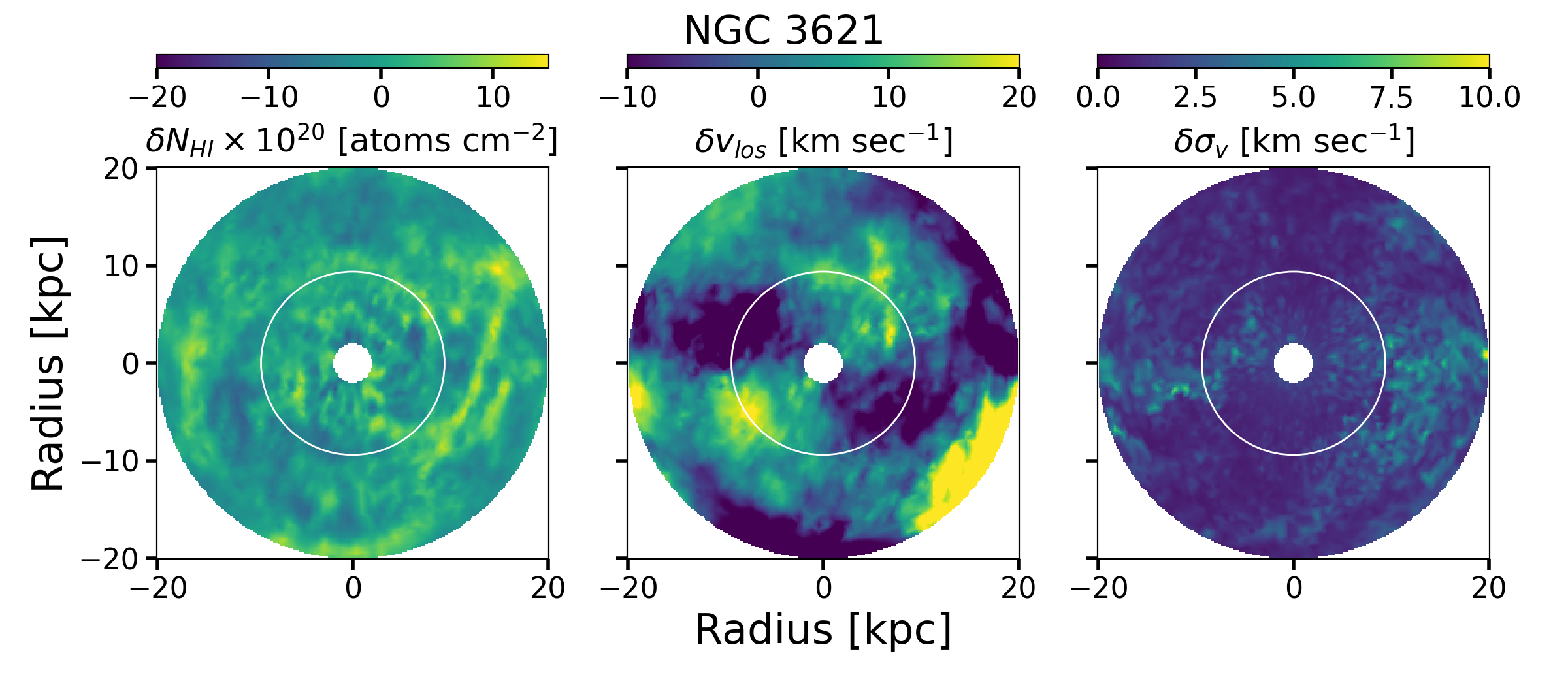}
\caption{The figure shows the fluctuations in column density ($\delta N_{HI}(R,\phi)$), line of sight velocity ($\delta v(R,\phi)$) and velocity dispersions ($\delta \sigma_v(R,\phi)$) for the galaxy NGC~3621 projected in the galactic plane. White circle represents the optical extent with radius $\sim R_{25}$.}
\label{fig:Corr3621}
\end{center}
\end{figure}

\subsection{Results}
\label{sec:results}

We first introduce the reader to our analysis with the galaxy \galaxy.  Distance to this galaxy is estimated to be  $6.6$ Mpc and the average inclination angle is $65^{\circ}$ \citep{2001ApJ...553...47F,deBlok+2008}. The moment maps from THINGS has an angular resolution of $16^{''} \times 10^{''}$, which corresponds to a linear scale of $510 \times 330 $ pc in the galaxy. 
%The scale height of this galaxy in \HI is not estimated in the literature. 
%\citet{10.1111/j.1365-2966.2004.07990.x} found that the  \HI scale height to scale length ratio of the spiral galaxies in their sample is $0.06  \pm 0.015$. 
 As discussed in section~\ref{sec:Meth}, considering the \HI extent $R_{HI}$  as an approximate estimate of  \HI scale length, we roughly estimate the  \HI scale height of this galaxy as $1.44$ kpc , which shows that $\Delta R <  3$ kpc. We choose $\Delta R = 570$ pc which is sufficiently larger than the beam size of $510$ pc. We choose $N_{\phi} = 144$ for all radii in this analysis.    % It is observed that the \ HI disk flares with radius \citep{1996AJ....112..457O,2016A&A...586A..98V}. }
%Note that, for the inner disc, the angular bins in azimuth may approach the linear resolution of the moment maps. However, for sufficiently small multipoles $m$, and hence rather larger linear scale corrugations, we should be able to properly sample them at smaller radii. 
We estimate the multipole decomposition of $f_0(R, \phi), f_1(R, \phi)$ and   $f_2(R, \phi)$ for multipoles up to $12$ for a range of $R$ values. The amplitude of the multipole moments beyond $m=12$ can not be measured with statistical significance. We estimate the fluctuations $ \delta N_{HI} (R, \phi), \delta v (R, \phi)$ and $\delta \sigma_v (R, \phi)$ using the method described in section \ref{sec:Meth}.

The top panel of Figure \ref{fig:sing_rad} shows different components of the line of sight velocity of \HI for \galaxy at a radius of $R=12$ kpc. The points with error bars correspond to the values of $f_{1}(R, \phi)$ as a function of azimuth $\phi$ with 5-$\sigma$ uncertainties. The black solid curve gives the first two moments ($m=0, 1$) of this function.  Bottom panel of the same figure shows $ \delta N_{HI} (R, \phi), \delta v (R, \phi)$ and $\delta \sigma_v (R, \phi)$ as a function of azimuth $\phi$. The estimated uncertainties in the values of each quantity are shown with a shaded region. Note that, the left y-axis represents the column density in units of $10^{20}$ atoms cm$^{-2}$, whereas the right y-axis represents the scale for velocity in km s$^{-1}$. We observe statistically significant oscillations in all three quantities here.  Figure~\ref{fig:Corr3621} show the fluctuations in column density ($\delta N_{HI}(R,\phi)$), line of sight velocity ($\delta v(R,\phi)$) and velocity dispersions ($\delta \sigma_v(R,\phi)$) for the galaxy NGC~3621 projected in the galactic plane. The white circle marks the $R_{25}$ for this galaxy.

\begin{figure}
\begin{center}
\includegraphics[scale=.5]{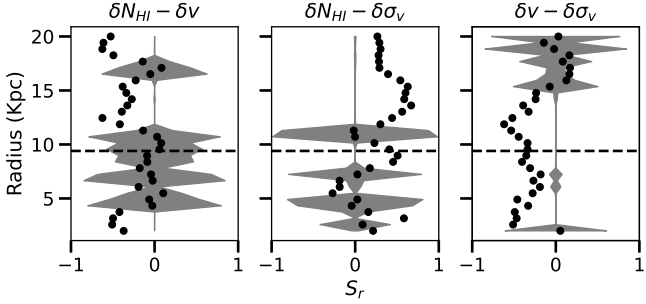}
\caption{The three columns show the Spearman rank correlation coefficients (black dots) estimated for the respective pairs of residual maps. Shaded region indicates chance correlation. The figures  share same y-axis giving the radius $R$. The horizontal black line represents $R_{25}$ of the galaxy.}
\label{fig:Corr_2d}
\end{center}
\end{figure}
\begin{figure}
\begin{center}
\includegraphics[scale=.28]{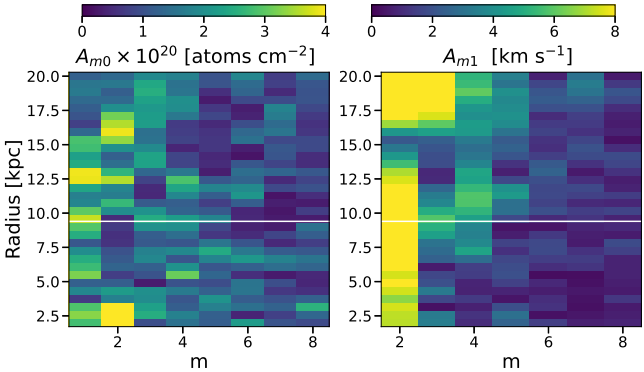}
\caption{Amplitudes of modes in residual maps of column density (left) and los velocity (right) are shown for different radii $R$. The white horizontal line corresponds to the $R_{25}$ indicating the optical disc of the galaxy.}
\label{fig:multi2d}
\end{center}
\end{figure}
\begingroup
\begin{table}
	\caption{ Table shows following parameters . Column 1: galaxy name; Column 2: \HI extent ($R_{HI}$), Column 3: average inclination angle (i); Column 4: Computed values of $h_z tan(i)$; Column 5: resolution beam size in pc units (\cite{Walter+2008}); Column 6: Choosen $\Delta R$ . References are already mentioned in table~\ref{tab:galpar} }
\setlength{\tabcolsep}{7pt} % Default value: 6pt
\renewcommand{\arraystretch}{.8}
\begin{tabular}{|l|c|c|c|c|c|c|l|}
\hline
	Galaxy   	& $R_{HI}$   & i    & $h_z tan(i) $  &Beam Size     &$\Delta R$      \\ \hline
		 &(kpc)      & (deg)  &  (pc)      &  (pc)       &   (pc)     \\ \hline
	NGC~2403 &  11.6     & 63     & 1370       &  130	 &  310 	\\ \hline
	NGC~2903 &  32.4     & 65     & 4170       &  658	 &  780  	 \\ \hline
	NGC~3621 &  24.0     & 65     & 3090       &  510	 &  570  	 \\ \hline
	NGC~4826 &  24.0     & 65     & 3090       &  440	 &  540 	  \\ \hline
	NGC~5055 &  44.1     & 59     & 4400       &  490	 &  590  		  \\ \hline
	NGC~6946 &  30.      & 33     & 1170       &  210	 &  430 	 \\ \hline
\end{tabular}
\label{tab:respar}
\end{table}
\endgroup

%As mentioned in section \ref{sec:Meth}, we do a multipole decomposition of the functions $f_i(R, \phi)$ at various radius. 

%Figure~\ref{fig:6946_Corro} show the estimated values of $ \delta N_{HI} (R, \phi), \delta v (R, \phi)$ and $\delta \sigma_v (R, \phi)$  for all radius in density plot. The horizontal line corresponds to $r_{25}$ of the galaxy indicating  the optical scale length. We observe interesting  correlations between the line of sight velocity and its dispersion. 
The Spearman rank correlation coefficient between the azimuthal values of all three pairs of $ \delta N_{HI} (R, \phi), \delta v (R, \phi)$ and $\delta \sigma_v (R, \phi)$  are plotted with black dots in the  Figure~\ref{fig:Corr_2d} for all radii. The shaded region represents the chance correlation between the pair of quantities, i.e, if the black dot lies outside the shaded region we can consider that the correlation is significant. The horizontal dashed line corresponds to the $R_{25}$ of the disc. 
We observe that the chance correlation is significantly higher when the correlation coefficients are near zero, ruling out any statistically significant correlation between the pair of variables. In some cases, we see that the pairs of variables stay correlated for successive radii, indicating that the correlation may be extended in the radial direction, whereas the sudden change in correlation in successive radii signifies local correlation. We observe that the line of sight velocity fluctuation $\delta v (R, \phi)$ is not correlated with its dispersion $ \sigma_{v} (R, \phi)$ indicating that the fluctuations seen in the line of sight velocity are not a result of moment map weighting. The strong anti-correlation seen between $ \delta N_{HI} (R, \phi)$ and $\delta v (R, \phi)$  at $R>10$ kpc strongly indicates towards the presence of disc corrugation. %The velocity dispersion, however show a correlation with the column density fluctuation which may have resulted from an additional  influence of self gravity in the vertical disc structure. 

\begin{figure}
\begin{center}
\includegraphics[scale=.37]{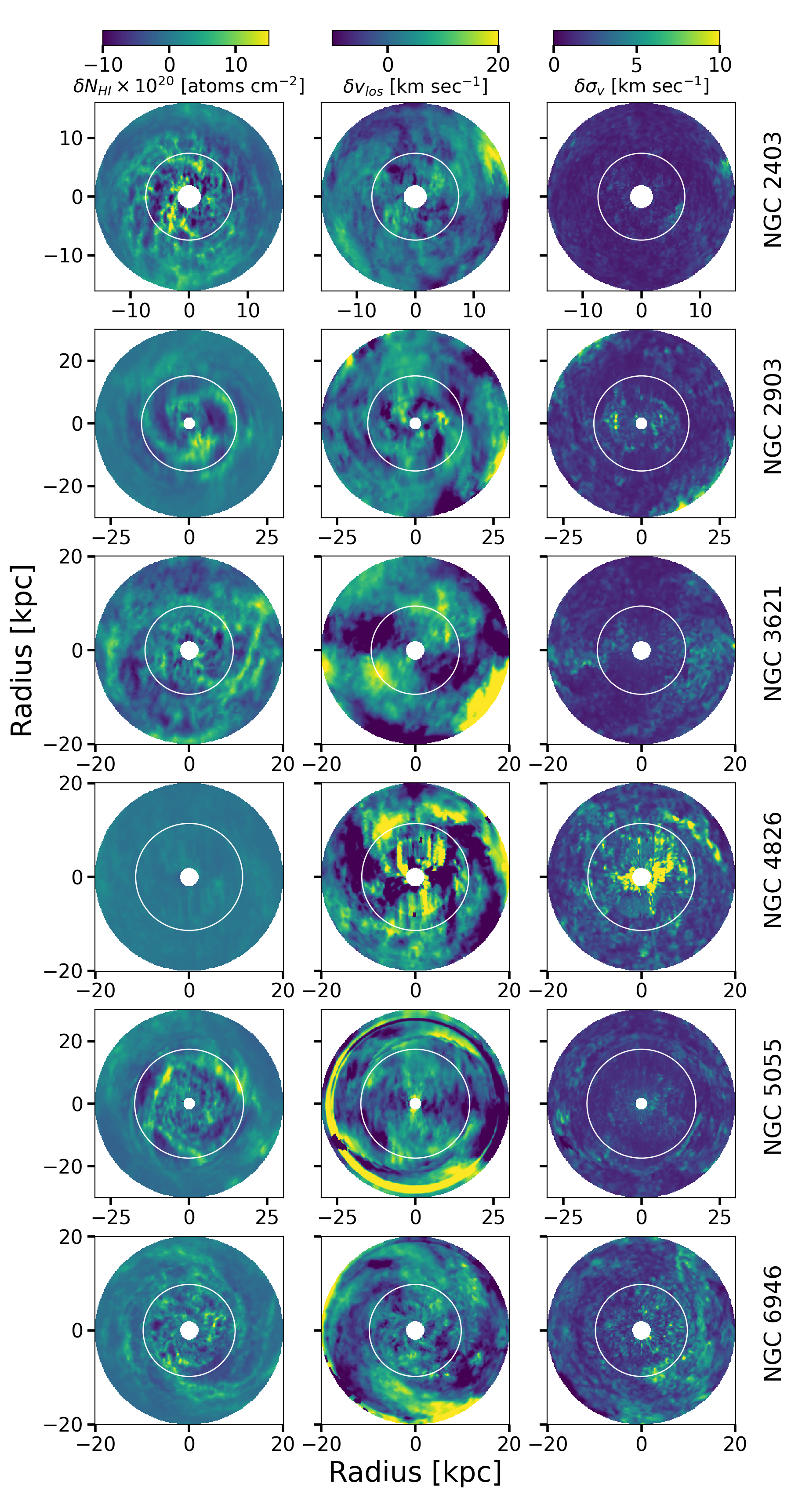}
\caption{Figure shows the fluctuations in column density ($\delta N_{HI}$), line of sight velocity ($\delta v$) and velocity dispersions ($\delta \sigma_v$) for all the six galaxies in our sample. White circle represents the optical extent of the respective galaxy with a radius $\sim R_{25}$. }
\label{fig:Corrall}
\end{center}
\end{figure}
\begin{figure}
\begin{center}
\includegraphics[scale=.44]{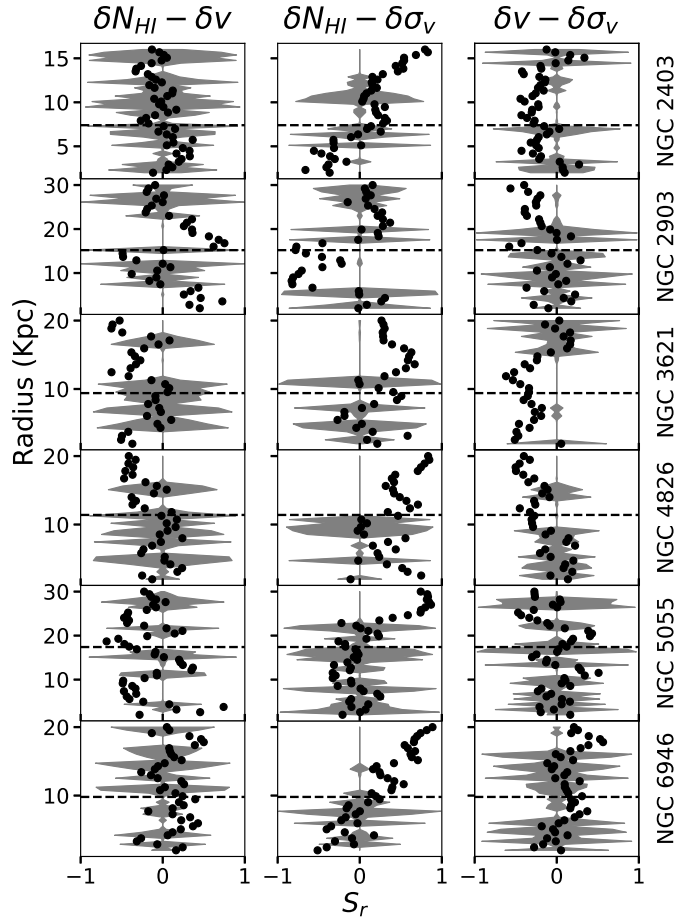}
\caption{The Spearman rank correlation coefficients for all three pairs of $ \delta N_{HI} (R, \phi), \delta v (R, \phi)$ and $\delta \sigma_v (R, \phi)$ as a function of radii for all the galaxies in our sample. The shaded regions corresponds to chance correlation. The horizontal lines denote $R_{25}$ for those galaxies.}
\label{fig:spearall}
\end{center}
\end{figure}

The amplitudes of different $m$ modes $A_{mj}$ are shown in Figure~\ref{fig:multi2d} for $m>0$ and $m>1$ for the column density (left, $j=0$) and line of sight velocity (right, $j=1$) respectively for different radii. We show for $m\leq8$ modes only, at higher $m$ the moment amplitudes are consistent with zero.  Clearly, only first few multipoles are having significant amplitudes. The horizontal lines in both panels mark  $R_{25}$ of the galaxy.  The distinct $A_{21}$ mode seen here can also be observed in Figure~\ref{fig:Corr3621}.
%Though we see that the spearman rank is often higher than its value that may come from chance correlation, the correlation coefficient itself is $\sim< 0.5$, indicating rather low correlation. Right panel of Figure~\ref{fig:Corr_2d} shows the cross correlation $\xi_{AB}(R, \psi)$ as a function of  lag in azimuth $\phi$ between all three pairs of $ \delta N_{HI} (R, \phi), \delta v (R, \phi)$ and $\delta \sigma_v (R, \phi)$ at different radius. The horizontal lines in both panel marks  $r_{25}$ of the galaxy as earlier. 

%\begin{figure}
%\begin{center}
%\includegraphics[scale=.3]{multi1d_3621.eps}
%\caption{Radial variation of the amplitudes $A_{m0}$ (top panel) and $A_{m1}$ (bottom panel) for first few multipoles are plotted with $3\sigma$ error bars. The vertical line corresponds to the $r_{25}$ of the galaxy. }
%\label{fig:multi1d}
%\end{center}
%\end{figure}

\begin{figure}
\begin{center}
\includegraphics[scale=.5]{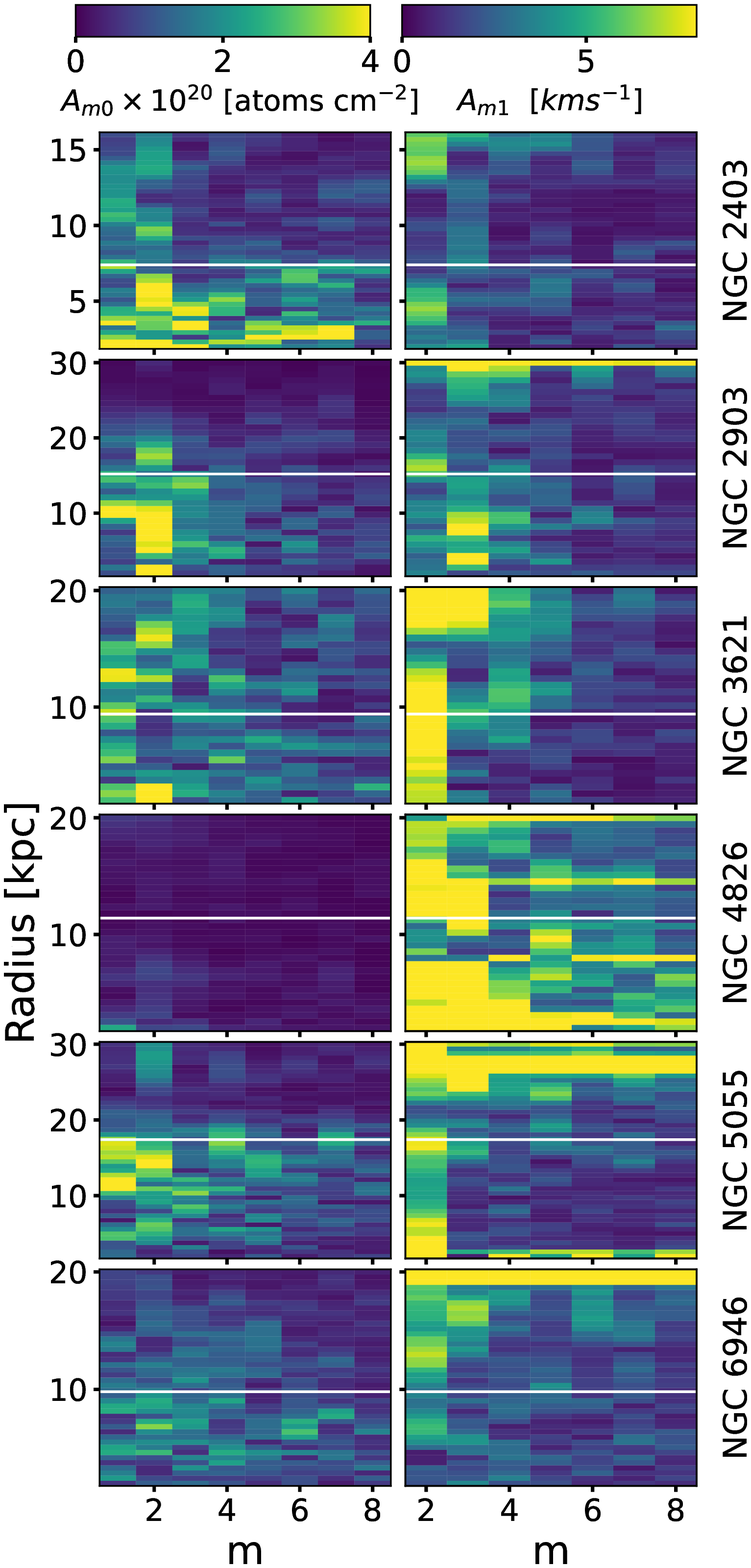}
\caption{The multipole amplitudes for residual maps of column density (left) and los velocity (right) at different radii for all galaxies in our sample.  The horizontal lines denote $R_{25}$ for those galaxies.}
\label{fig:multiall}
\end{center}
\end{figure}

\begin{figure}
\includegraphics[width=0.5\textwidth]{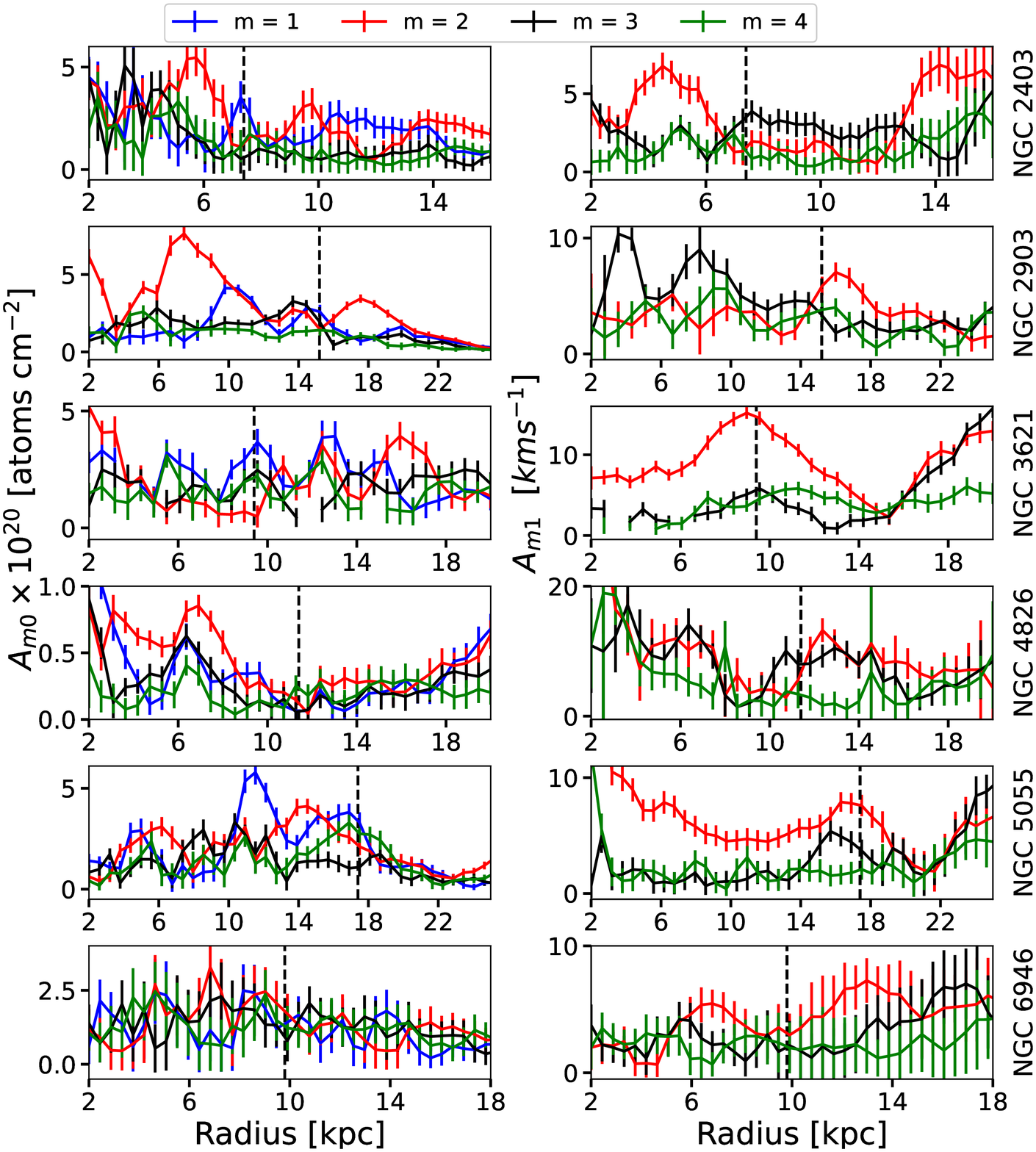}
\caption{First few  multipole amplitudes $A_{m0}$(left) and $A_{m1}$ (right) at different radius for all galaxies in our sample. The vertical lines denote $R_{25}$ for those galaxies.}
\label{fig:multiall1d}
\end{figure}

%\begin{figure}
%\begin{center}
%\includegraphics[scale=.55]{crosscor_all.eps}
%\caption{The  cross-correlation function  for all three pairs of $ \delta N_{HI} (R, \phi), \delta v (R, \phi)$ and $\delta \sigma_v (R, \phi)$ for different radius for all galaxies in our sample.  The horizontal lines denote $r_{25}$ for each galaxies.}
%\label{fig:crossall}
%\end{center}
%\end{figure}

%We plot the first four significant multipole amplitudes  of column density (top) and first three significant multipoles for the  line of sight velocity (bottom)  in Figure~\ref{fig:multi1d} with error bars showing three-sigma uncertainties. The vertical dashed line indicates $r_{25}$ of the galaxy.

 We perform the above analysis for all the six galaxies in our sample.  Table~\ref{tab:respar} shows the $R_{HI}$, $i$, $h_z\, \tan\, i$, the resolution beam size in $pc$ units and the chosen value for $\Delta R$ for all the galaxies. Note that $\Delta R $ is at least $5$ times smaller than $h_z\, tan\, i$ and in all cases greater than the beam size. Given that the \HI disc flaring is observed to be not more than a few times in the inner and outer parts of the disc \cite {2008ApJ...685..254B}, our choice of $\Delta R$ suffices. 
 Figure~\ref{fig:Corrall} shows the fluctuations in column density ($\delta N_{HI}(R,\phi)$), line of sight velocity ($\delta v(R,\phi)$) and velocity dispersions ($\delta \sigma_v(R,\phi)$) for all the six galaxies in our sample.
We show the Spearman rank correlation coefficients for the pairs of $ \delta v (R, \phi)$, $\delta \sigma_v (R, \phi)$ and $\delta N_{HI} (R, \phi),$for all the galaxies in Figure~\ref{fig:spearall}.  Each row represents a galaxy from our sample. The radial \HI extent of each galaxy is different and is shown on the y-axis. The dashed horizontal line corresponds to the $R_{25}$ of the galaxy. We note that the line of sight velocity and its dispersion mostly do not show any significant correlation indicating the absence of influence from moment 0 map weighting. All galaxies (except NGC~2903) show a correlation between column density and line of sight velocity dispersion at large radii. There is also a significant correlation in the column density and line of sight velocity for NGC~5055, NGC~4826, NGC~3621 and  NGC~2903 at various radii. These two correlations suggest the presence of vertical bending modes in these galaxies.

The multipole amplitudes $A_{m0}$ for all six galaxies are plotted for $m>0$ in the left-hand side of Figure~\ref{fig:multiall}, the right-hand panel of the same figure corresponds to multipole amplitudes $A_{m1}$ with  $m>1$. Amplitudes for $m>8$ are not shown as they are found to be negligible in our sample. The horizontal white dashed line corresponds to the $R_{25}$ of each galaxy. All the lower modes show strong radial dependence showing maximum amplitudes only at certain radii. 
%Most of the galaxies seem to have oscillatory pattern in column density and line of sight velocity fluctuation at lower multipoles. 
The case of NGC~4826 is particularly interesting, as it does not show much variation in column density residual but shows significant peaks in velocity residual. To observe the variation of the multipole amplitudes as a function of radius we show the first few significant multipole amplitudes for all the galaxies in  Figure~\ref{fig:multiall1d}  with three sigma uncertainties. As can be seen, the observed oscillations have large observational significance in most of the cases. 
%Henceforth we consider all the statistically significant modes in this figure as arising from bending modes in the disc.

\subsection{Identifying the modes}
\label{sec:ident}

\begin{figure*}

\begin{subfigure}{}
\includegraphics[width=0.49\textwidth]{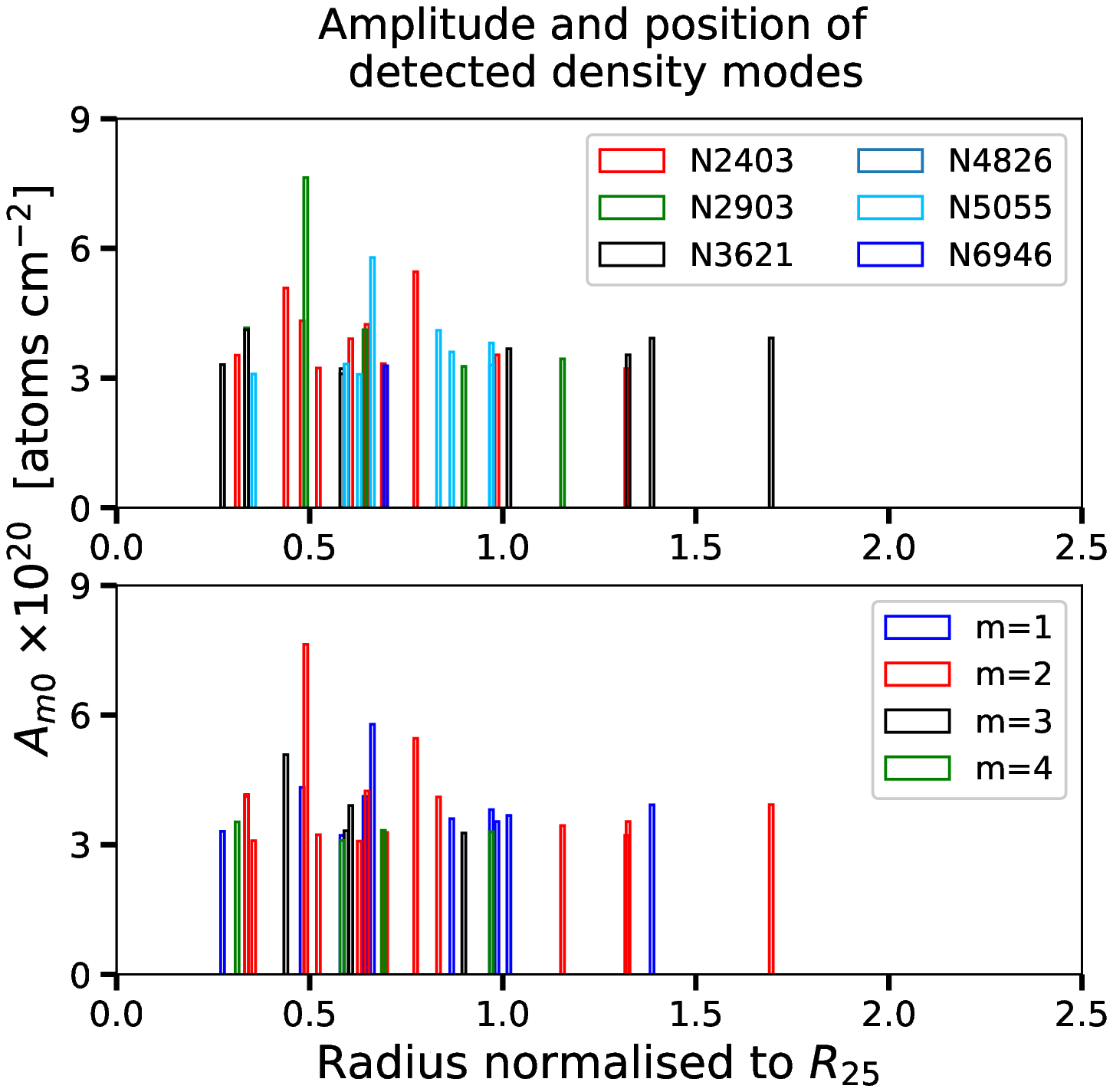}
%\subcaption{\bf (a)}
\end{subfigure}
\begin{subfigure}{}
\includegraphics[width=0.49\textwidth]{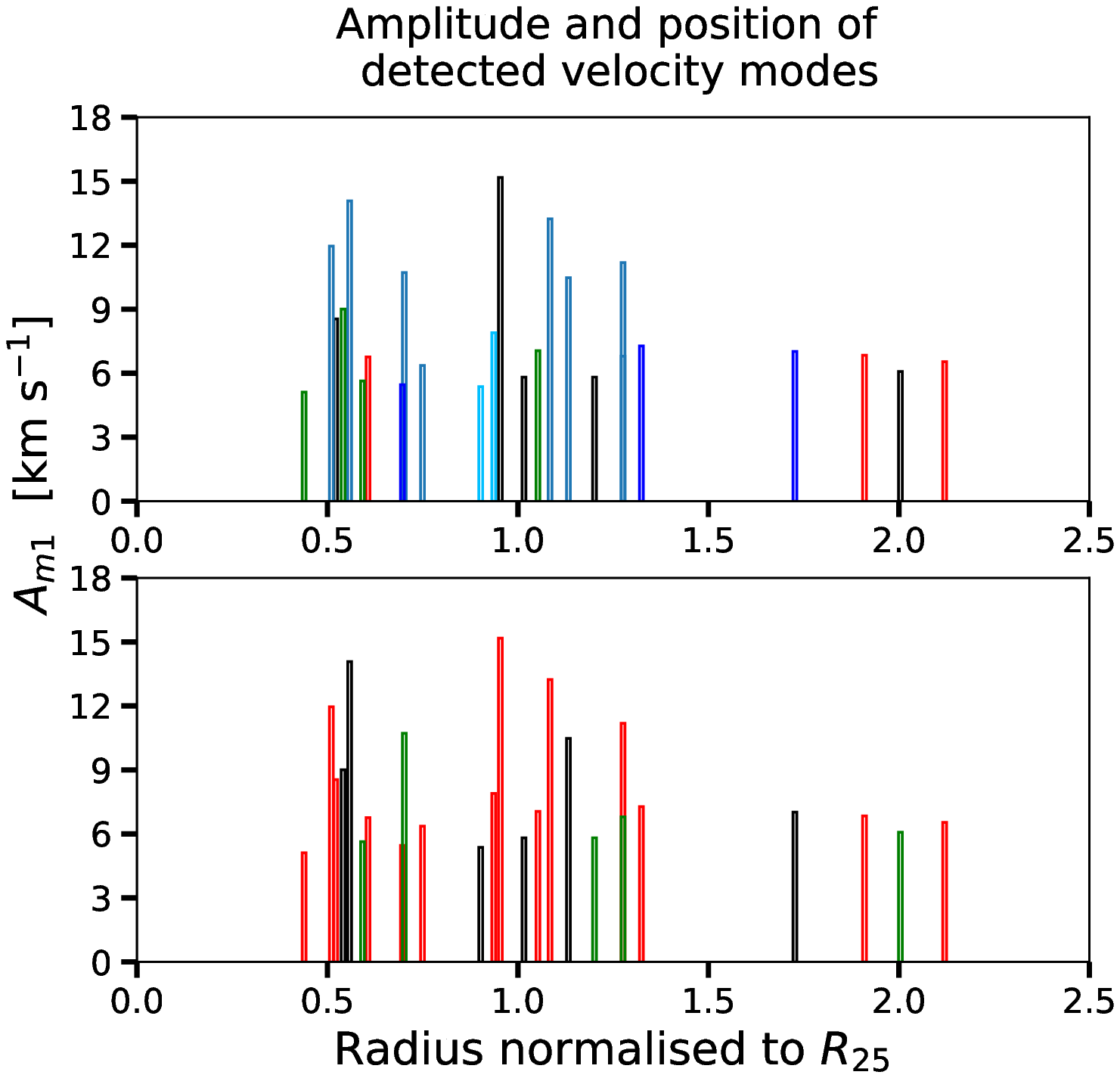}
\end{subfigure}\\
%{\bf (a)\hspace{9cm}(b)} \\
\caption{Column density (left) and line of sight velocity (right) mode peaks shown for all galaxies.
The vertical axes show their maximum amplitude and the horizontal axes show the radial positions in their host galaxies. For the ease of comparing, radial positions are normalized by $R_{25}$ of respective galaxies. In the top panel, the amplitude of all modes for different galaxies (color coded) are shown. The bottom panel shows the same color coded mode-wise. Note that we do not  include m=1 velocity mode here. 
}
\label{fig:Fig8}
\end{figure*}

\begin{figure*}
\begin{subfigure}{}
\includegraphics[width=0.49\textwidth]{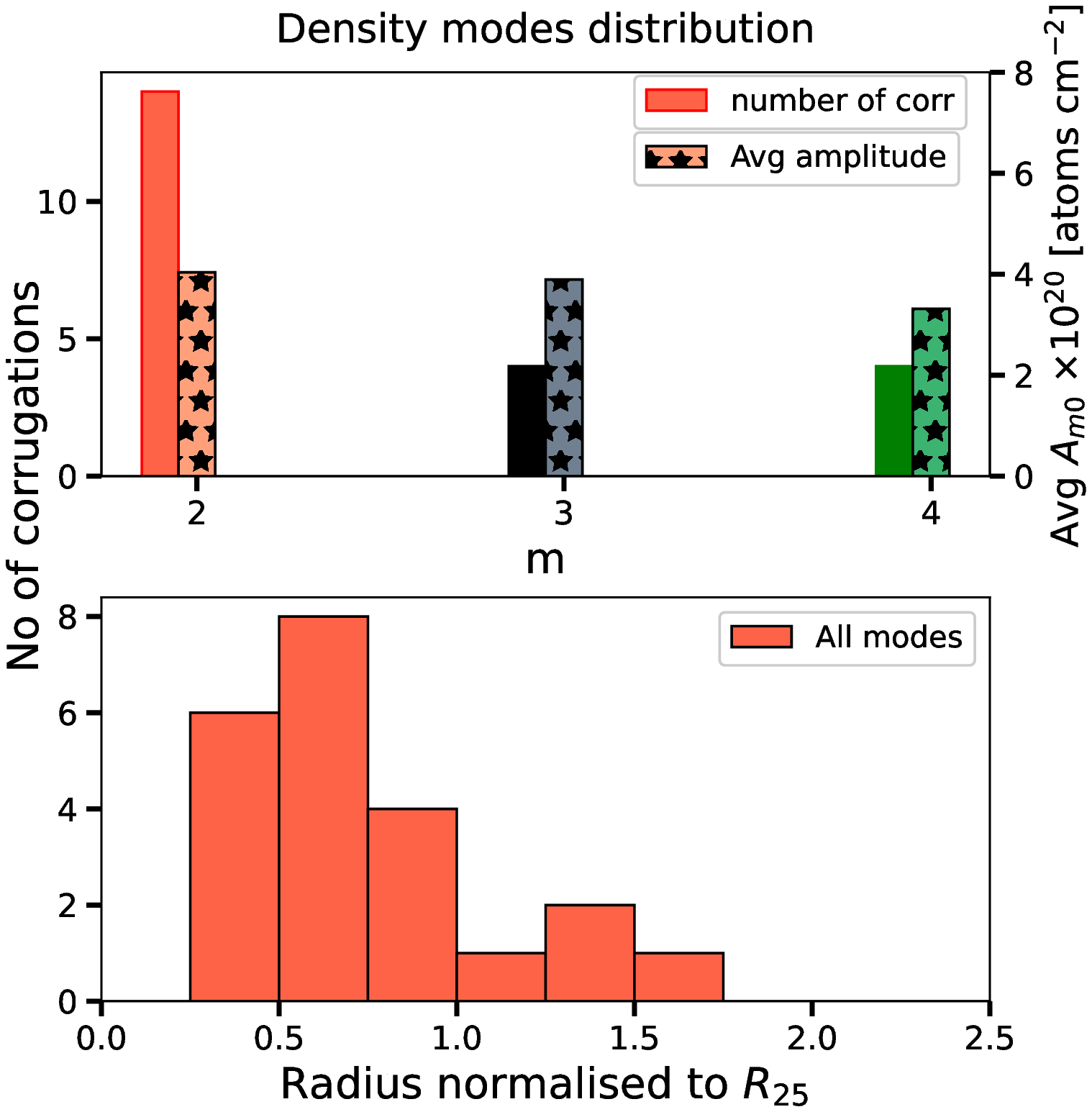}
\end{subfigure}
\begin{subfigure}{}
\includegraphics[width=0.49\textwidth]{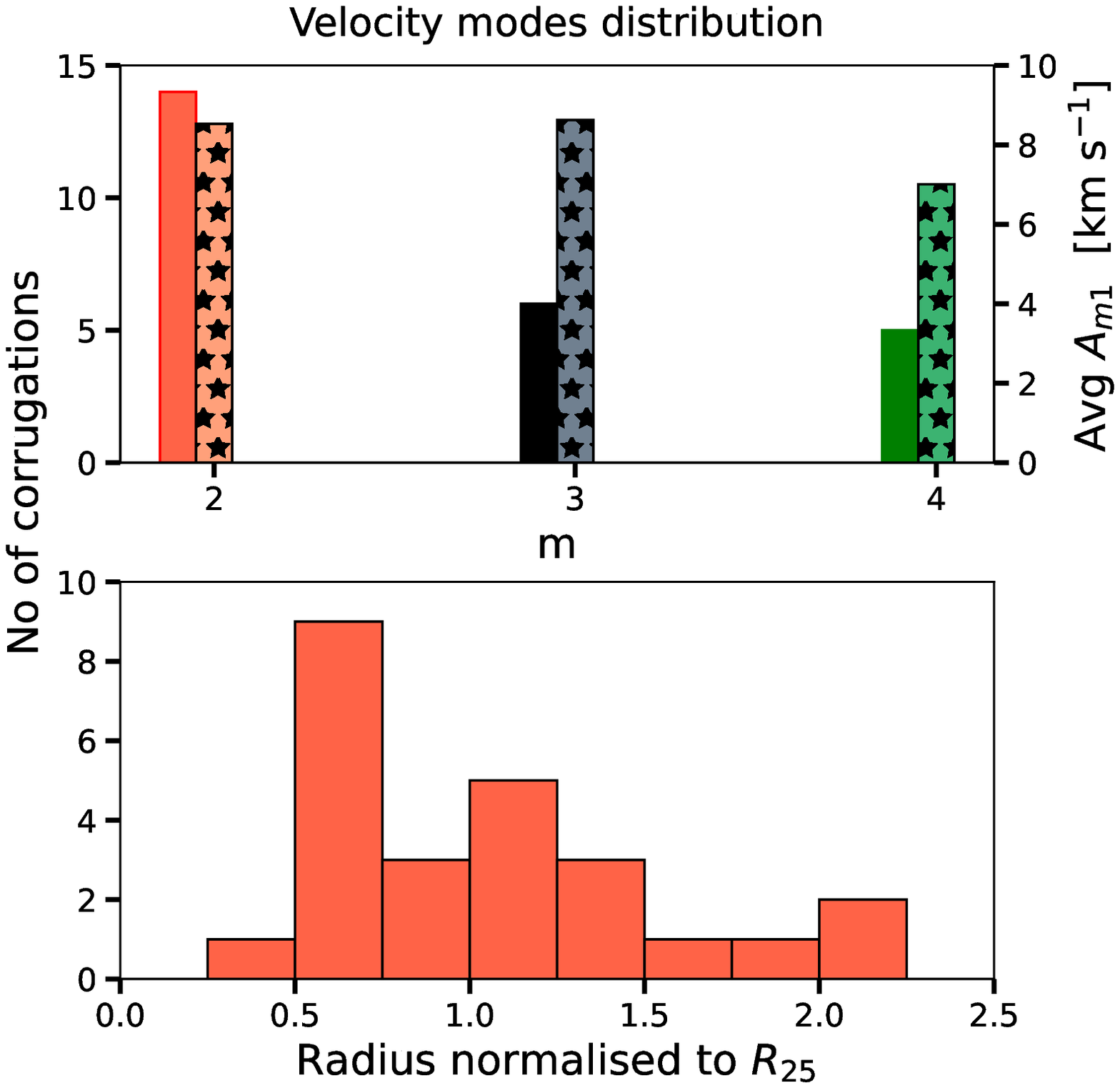}
%{\bf (b)}
\end{subfigure} \\
%{\bf (c)\hspace{9cm}(d)} \\
\vspace{4pt}
%\caption{ }
\caption{Various statistics related to the column density (left) and velocity modes (right) are shown. Top panels show the number of detected density and velocity peaks for m = 2, 3 \& 4 (solid colors)  alongside with their average amplitudes (bars with stars) considering all galaxies. The bottom panel shows a histogram of radial position of these modes. }
\label{fig:Fig9}
\end{figure*}

In the right panel of Fig \ref{fig:multiall1d} we show the radial variation of the amplitude of the modes = 2,3 and 4 obtained from the \HI residual velocity map for all six galaxies. 
 As m = 0 component of velocity overlaps with the galaxy's systemic radial velocity and m = 1 component overlaps with the galaxy's rotation velocity, they are not shown here.
Amplitudes for modes m $> 4$ are not shown as they are negligible (see Fig \ref{fig:multiall}). In fact, NGC 4826 is the only galaxy that shows a peak of above 5 km s$^{-1}$ in m = 5.

Here we describe the method we use to identify the modes with an example. Consider the amplitude of  m = 2 velocity mode (red line in Fig \ref{fig:multiall1d}) in NGC~3621. This mode has a significant presence ($> 8$ kms$^{-1}$) in the galactocentric region of $6-12$ kpc and reaches a peak of 15 kms$^{-1}$ at 9 kpc. This kind of scenario is very similar to what are seen in simulations \citep{Widrow+2012, Carlin+2013, Gomez+2013, Gomez+2017} where the maximum bending is localised in R. As predicted in theory and witnessed in simulations \citep{BT21}, these `radial peaks' move outward with time. 
 Similar but smaller crests in m = 3 \& 4 of amplitude 6 kms$^{-1}$ are seen at 10 and 11 kpc respectively.
We regard this continuous significant amplitude over a radial range with a crest in between as the sign of a bending mode.
The amplitudes of m = 2 \& 3 also show a steep increase towards the limit of the observed \HI disc but we do not interpret these as bending modes as the significance of the detection is much less there. Thus we find 3 modes in NGC 3621: $m=2$ with a peak of 15 kms$^{-1}$ at 9kpc; $m=3$ with a peak of 6 kms$^{-1}$ at 10 kpc and $m=4$ with a peak of 6 kms$^{-1}$ at 11 kpc.

 Similarly, other galaxies in our sample also show these radial peaks spread across their entire discs. 
 Since the location of a peak and its amplitude can vary with time  \citep{Widrow+2012, Carlin+2013, Gomez+2013, Gomez+2017, Poggio+2021}, the observed snapshot of the mode spectrum in a galaxy may be useful in predicting the stage of tidal interaction the disc is in.  We perform an exercise of identifying the peaks in both column density and velocity, their corresponding amplitudes and locations for all the sample galaxies.  
 We find that in column density there are a total of 32 peaks for all multipoles (m=1,2,3 and 4) with amplitude above 3 $\times 10^{20}$  atoms cm$^{ -2}$. There was no density peak observed for the galaxy NGC~4826. For the line of sight velocity, we find 25 peaks in total for multipoles  2,3 and 4 with amplitude above 5 km s$^{-1}$. In both cases, we restrict ourselves in radius between 0.4 R$_{25}$ and the observed outer limit of the discs. We do not consider the inner disc region because there the self-gravity is strong enough to resist the formation of bending waves \citep{Saha&Jog2006}.
 Furthermore, the stellar spiral structure may also contribute to coherent variations in los velocity of \HI and hence not all peaks in this region may be attributed to bending waves. \cite{Alfaro+2001} and \cite{Sanchez+2015} find a close association between velocity peaks and emission peaks in H$_{\alpha}$ in the inner disc regions. 
 %We also do not consider the large  amplitudes at the edges of the \HI discs  as peaks for reasons already mentioned above.
 Figure~\ref{fig:Fig8} summarises the result of this analysis. The plots in the left column correspond to the column density and that are in the right corresponds to the line of sight velocity. The panels show the amplitude of all the identified density/velocity modes colour coded galaxy-wise (top panel) and mode-wise (bottom panel) at their corresponding radial location (scaled to the optical disc radius, R$_{25}$). A statistical analysis of these modes coming from all the galaxies together is shown in Figure~\ref{fig:Fig9}. The top panel of this figure gives the number of appearances of different modes (left vertical axis) and their average amplitudes (right vertical axis) for m=2,3 and 4 in column density and velocity.  The bottom panel shows how all the modes are radially distributed.  We do not include the m=1 column density mode here such that comparison between the velocity and density modes is easier.

\subsection{Preliminary inferences}
\label{sec:infer}

 Although we have analysed only six galaxies, we are surprised to find as many as 25 radially unique velocity peaks in the mode window of 2-4. NGC~2403 and NGC~6946 show a minimum of three peaks each whereas NGC~4826 shows up to 7 peaks in velocity. It is important to note here that there is no galaxy in this sample that does not show any velocity modes at all indicating that all the galaxies may be definitely perturbed to an extent.  This is in accordance with recent simulations by \cite{Gomez+2017} who predict corrugations to be a very common phenomenon.  Whereas in the case of density, we identify around 32 peaks with modes ranging from m=1 to m=4, the majority of them have amplitudes of $~\sim 3$ $ \times 10^{20}$  atoms cm$^{-2}$. NGC~4826 is an interesting galaxy in our sample, where not a single peak is detected in density space, while in velocity space it has the largest number of peaks among the sample. A single-density peak is found near the optical extent in NGC~6946. The remaining galaxies have a significant number of density peaks ($> 5$ peaks) identified over different radial positions with NGC~2903 having the highest amplitude density peak ($\sim 7.6$ $ \times 10^{20}$ atoms cm$^{-2}$).  We find the following interesting trends based on the results presented in Figure~\ref{fig:Fig8} and Figure~\ref{fig:Fig9}:   
%In Fig \ref{fig:mdist} we show some simple histograms of the distribution of the identified modes. The top panel shows how the three modes (seen in different colours) are represented in their number count (plain bars) and their average amplitude (starred bars), whereas the bottom panel shows how they are radial distributed. 

\begin{enumerate}
    \item More than 50\% of the velocity peaks belongs to m = 2 modes while about 20\% - 25\% belong to m = 2 and 3.  This indicates that m=2 may be easier to excite or it is more long-lived compared to the other two. This indication is in agreement with the simulations of \cite{Poggio+2021} where they find that m = 2 is more stable than the other two modes studied (m = 0 and 1). As they have not followed m $> 2$ modes, we cannot directly compare our results with theirs in this aspect.  Around 75\% of the density peaks that are identified are either mode m= 1 or 2.   
    \item the average amplitude in each velocity mode is about 8 km s$^{-1}$  and density mode is $\sim$4 $ \times 10^{20}$ atoms cm$^{ -2}$ 
    \item The maximum number of density modes are seen in the inner galaxy. This is expected if these are indeed spiral arms.
    \item The velocity peak distribution shows a bi-modal behaviour. The first maximum overlaps with the density mode maximum $\sim$ 0.6 R$_{25}$ suggesting that these velocity peaks may be related to spiral arms and not due to bending waves. The second smaller maximum occurs at the optical edge and are more likely due to bending waves. 
%    \item the region around the edge of the optical disc (0.5 R$_{25} <$ R $<$ 1.5 R$_{25}$) seems special from two aspects - most of the detected peaks are found here and the largest amplitude peaks are also found here, in both velocity and density space.
    \item  We observe in  figure~\ref{fig:multiall}, that for all galaxies except NGC~4826, at least one peak in density and velocity space is located in the same position. These can be originating from the in disk motion of the gas in an undulated disk as discussed with  figure~\ref{fig:toymodel}.
    \item Since we do not find any associated column density mode for any of the velocity modes for NGC~4826, it is quite likely that for this galaxy, the observed velocity modes are arising from a classical vertical corrugation.
%       \item  From the radial distribution of the modes in the bottom panel ol figure~\ref{fig:Fig9}, it is clear that a major fraction of the detected density and velocity modes (m=2, 3 and 4)  are located around $\sim 0.6R_{25} $. In addition, there is a secondary peak at $R_{25}$ for the velocity histogram.
\end{enumerate} 

The above points raise intriguing questions like a) what makes m = 2 more stable than other modes? b) why are the corrugations in \HI disc more concentrated around the edge of the optical disc rather than the edge of the \HI disc? c) similarly, why are the maximum amplitudes seen at the optical edge rather than the \HI disc edge? These questions can be addressed in two possible ways - via simulations that include an extended \HI disc component; and by identifying velocity corrugations in a statistically large sample of galaxies (in both optical and \HI discs). We hope to see these studies in future.

\section{Discussion}
\label{sec:disc}

Interestingly, \cite{Poggio+2021} have studied the evolution of tidally generated bending wave modes in a Milky Way like disc throughout a complete interaction covering several plane-crossings of an Sgr like a satellite. From a distance of 300 kpc between the two at time t = 0, until the Sgr's complete merger 7 Gyrs later, the disc goes through several phases of perturbation and relaxation. Their simulations (see their Fig 5) show that the following happens simultaneously with subsequent plane-crossings.

\begin{enumerate}
    \item bending waves are generated first in the outer disc and then slowly spread inwards
    \item the perturbation begins with the lowest order mode (m = 0) and chronologically moves to higher orders
    \item the amplitude of bending waves increases with time
\end{enumerate}

All galaxies in our sample have smaller companions \citep{Carlsten+2022} and
these findings are extremely helpful in placing a galaxy in the early, mid or later stages of an interaction based on its observed bending wave spectrum. 
%But obtaining the amplitudes of the different bending modes from observations is the real challenge as discussed in Section \ref{sec:Intro}. However a number of simulations that have studied tidally generated vertical perturbation \citep{Widrow+2012,Carlin+2013,Gomez+2013,Gomez+2017} have established that the generation of bending waves invariably leads to coherent vertical velocity undulations, as seen in our sample and also in \cite{Gomez+2021}. 
Based on this, we can make a preliminary inference (from Figure~\ref{fig:multiall}) on the depth of the interaction level that each of our sample galaxies may be in. In NGC~2403, NGC~5055 and NGC~6946, the velocity peaks are mostly seen in the outer discs and are of low amplitudes, implying that they may be in the early stages of an interaction. NGC~2903 shows bending within the optical disc has all three modes but amplitudes are not high. This could mean that the galaxy is somewhere in the mid-stage of an interaction. In NGC~3621 and NGC~4826, we notice a larger number of peaks, spread across most of the disc and also with high amplitudes. Hence they may be in advanced stages of interaction.

\cite{Poggio+2021} also find a $90^{\circ}$ shift between the phase angles of the bending wave and the corresponding velocity corrugation for m = 2. While we cannot measure this for the lack of direct measurement of bending waves in our sample, we note that theoretically, the phase difference should be $\lambda/4 = \pi/2m$ where $\lambda = 2\pi R/m$ is the wavelength of the bending wave for any annulus of radius R (see Figure \ref{fig:toymodel}). This means the expected phase difference is $45^{\circ}$ for m = 2. This is puzzling and we hope to follow it up in future.
%\newpage

\section{Conclusions}
\label{sec:conc}

A number of recent simulations have confirmed the theoretical predictions of bending waves being spontaneously generated during a tidal interaction. They have also shown that the generated bending waves invariably lead to an exactly similar pattern in the disc's vertical velocity, but with a phase difference. Because of their tidal origin, the bending waves are also expected to be highly common \citep{Gomez+2017, ChequersWidrow2017} but are seldom noticed owing to their low amplitudes and unfavourable disc inclinations. So, looking for their kinematic imprint appears to be far more feasible and may be equally resourceful, in understanding the bending waves and through them the evolution of disc galaxies via low-mass tidal interactions.

In this paper, we look for the kinematic bending modes in the \HI velocity maps of six galaxies from the THINGS sample: NGC 2403, NGC 2903, NGC 3601, NGC 4826, NGC 5055 and NGC 6946. We subject the residual velocity maps and column density maps (after subtracting local average components) to multipole analysis (performed on each annular ring centred around the galaxy) and find several velocity and density peaks belonging to different modes, with varying amplitudes, spread across the entire \HI disc. We are unable to access the kinematic m = 0 and 1 as they are mixed with the disc's linear and rotational velocity components respectively. We are able to filter out about 25 velocity peaks with amplitude above 5 km s$^{-1}$ and lying between 0.4 R$_{25}$ and the disc's reliable outer limit. The maximum amplitude detected is 15 km s$^{-1}$. About 32 density peaks are detected above 3 $ \times 10^{20}$ atoms cm$^{ -2}$  in the similar radius range with a maximum amplitude of 7.6 $ \times 10^{20}$ atoms cm$^{ -2}$ . We find the following interesting trends from the detected modes (belonging to m = 2,3 and 4 only) in this limited sample study:

\begin{enumerate}
    \item All of the sample galaxies appear to be vertically perturbed
    \item More than 50\% of the kinematic modes are m = 2, suggesting that they may be more long-lived than the higher-order modes.
    \item The average amplitude in each of m = 2,3,4 kinematic modes is the same -$ \sim$ 8 km s$^{-1}$
    \item Maximum number  of non-axisymmetries in column density as well as los velocity are seen $ \sim $0.6 R$_{25}$ suggesting spiral arms as their likely cause.
    \item A second smaller maxima in velocity non-axisymmetries is seen at the optical edge suggesting bending wave origin. The amplitude of the modes are also larger in this region.
%    \item  Both the density and velocity peaks seem to be concentrated around the optical disc edge. 
%    \item the amplitude of the modes are larger near the optical edge. 
%    \item the modes seen within the optical disk are most likely originating from an undulating disk with the tangential velocity following the disk undulation. The modes seen near and outside the disk are most likely from classical vertical velocity corrugation. 
\end{enumerate}

 It is important to note that several physical effects including bending waves, coherent non-axisymmetric and/or vertical velocity components and density waves in the disk can give rise to the observed column density and line of sight velocity modes. We believe that the observed column density and velocity modes outside the optical disc are primarily from bending waves. A detailed study of the observational signatures of these different effects is needed to establish the origin of the modes.
We urge for similar studies with a larger sample and also hope to see simulations including the gas component.

\section*{Acknowledgement}
We thank Jayaram Chengalur for initiating this work. We thank Ralf-Juergen Dettmar, Peter Kamphuis, Nissim Kanekar and Kanak Saha for useful discussions. MN acknowledges Department of Science and Technology - Innovation in Science Pursuit for Inspired Research (DST-INSPIRE) fellowship for funding this work. CN acknowledges post-doc fellowship support from National Centre for Radio Astrophysics, Pune. PD acknowledges DST-INSPIRE faculty fellowship support for this work. Authors thank the anonymous referee for suggestions that have improved the presentation of the paper significantly. 
\section*{DATA AVAILABILITY}
The \HI moment maps used in this analysis are available on the THINGS website \\ ( \href{https://www2.mpia-hd.mpg.de/THINGS/Data.html}{https://www2.mpia-hd.mpg.de/THINGS/Data.html}).

\bibliographystyle{mn2e}
\bibliography{references}

\end{document}